\documentclass[a4paper,11pt]{article}
\usepackage{pos} 

\definecolor{hu-berlin-blue}{RGB}{0,65,137} 

\title{The complex potential from 2+1 flavor QCD using HTL inspired approach}

\author*[a]{Dibyendu Bala}
\author[a]{Olaf Kaczmarek}
\author[b]{Rasmus Larsen}
\author[c]{Swagato Mukherjee}
\author[b]{Gaurang Parkar}
\author[c]{Peter Petreczky}
\author[b]{Alexander Rothkopf}
\author[d]{Johannes Heinrich Weber}

\affiliation[a]{Fakult\"at f\"ur Physik, Universit\"at Bielefeld, D-33615 Bielefeld, Germany}

\affiliation[b]{Faculty of Science and Technology, University of Stavanger, NO-4036 Stavanger, Norway}

\affiliation[c]{Physics Department, Brookhaven National Laboratory, Upton, New York 11973, USA}
\affiliation[d]{Institut f\"ur Physik \& IRIS Adlershof, Humboldt-Universit\"at zu Berlin, D-12489 Berlin, Germany}

\abstract{We have studied finite temperature complex static quark-antiquark potentials for 2+1 flavor QCD using highly improved staggered action with physical strange quark masses and light quark masses corresponding to a pion mass of 161 MeV. We calculated the potential using  Wilson line correlators fixed in Coulomb gauge. For the extraction, we have used HTL motivated parametrization of the correlators. We found that the real part of the potential is screened above the crossover temperature and it's close to singlet free energies, whereas the imaginary part is increasing with both distance and temperature.}

\FullConference{%
 The 38th International Symposium on Lattice Field Theory, LATTICE2021
  26th-30th July, 2021
  Zoom/Gather@Massachusetts Institute of Technology
}

\begin{document}
\maketitle

\section{Introduction}
Quarkonia, the bound states of heavy quark-antiquark pairs have been played an important role for the understanding of the Quark-Gluon Plasma (QGP), particularly after the proposition by Matsui and Satz \cite{MS} that the suppression of quarkonium can be an important signal of the formation of QGP. One way to understand the propagation of quarkonia inside the plasma is by defining a thermal static quark-antiquark potential \cite{LPRT,BGVP}. The potential is defined by the following expression,
\begin{equation}
	V(r,T)=i \lim_{t\rightarrow \infty} \frac{\partial \log W_{T}(r,t)}{\partial t}.
	\label{defn}
\end{equation}

Here $W_{T}(r,t)$ is a thermal averaged Wilson loop in real-time. The potential is only defined if the above limit exists. At zero temperature, the existence of a potential follows from the transfer matrix argument. At finite temperature, however, the existence of the above limit is a non-trivial condition. In leading order Hard Thermal Loop (HTL) perturbation theory it has been found that a potential can be defined and in contrast to zero temperature the finite temperature potential is complex \cite{LPRT}. The real and imaginary part of the potential for $r\sim {1}/{m_D}$ is given by ,
\begin{equation}
\begin{split}
V_{re}(r)  =  - \frac{g^2 C_F}{4 \pi} \left(\frac{e^{- m_{D} r}}{r} + m_{D}\right) \\
V_{im}(r) = \frac{g^2 C_F}{4\pi} T \int\limits_0^\infty dz \,
\frac{2 z}{\left(z^2+1\right)^2} \left[ 1 - \frac{\sin z m_D r}{z m_D r} \right].
\end{split}
\label{HTL_pot}
\end{equation}
Here $g$ is the QCD coupling constant at scale $T$ and $C_F=4/3$ for QCD.
The real part is the Debye screened version of Coulomb potential with a Debye mass $m_D$, whereas the imaginary part is approaching zero at short distances and saturates at a long distance.

The definition of the potential in Eq.~(\ref{defn}) requires a real time Wilson loop, but on the lattice we can only calculate Wilson loops $W_{T}(r,\tau)$ in imaginary time $\tau \in (0,\beta=1/T)$. The real time Wilson loop can in principle be obtained from imaginary time Wilson loop $W_{T}(r,\tau)$ by analytic continuation, which is done by the spectral function $\rho(r,\omega)$ \cite{RHS} as follows,
\begin{equation}
\begin{split}
W_{T}(r,\tau)=\int d\omega \exp(-\omega\,\tau ) \rho(r,\omega)\\
W_{T}(r,t)=\int d\omega \exp(-i\,\omega\,t ) \rho(r,\omega).
\label{cont}
\end{split}
\end{equation}
The extraction of the spectral function $\rho(r,\omega)$ from a finite set of data of $W_{T}(r,\tau)$ is an unstable problem without any initial prior of the spectral function. As a result various spectral functions are possible which satisfy the lattice data. Some of the spectral functions, however, do not support any well defined limit of Eq.~(\ref{defn}) \cite{BKLMPPRJ}. For various applications, like the construction of the spectral function for point vector current correlator \cite{BMV} or in open quantum system studies of quarkonia \cite{AR,KAAR} one needs to have a potential. Because of this we will constrain our spectral function such that a potential exist, quantiatively we will consider spectral functions that satisfies
\begin{equation}
	\lim_{t\rightarrow \infty} \frac{\int \omega \rho(\omega,r)\exp(-i\omega\,t)\,d\omega}{\int \rho(\omega,r)\exp(-i\omega\,t)d\,\omega}=V_{re}(r,T)-i\,V_{im}(r,T).
\label{condition}
\end{equation}
The Bayesian reconstruction method in \cite{BKR} generates spectral functions that automatically satisfy the above condition. In this proceeding, we will use a HTL motivated parametrization \cite{BD} which also satisfies the above condition.
\section{Method}
In HTL perturbation theory, the leading order result for the Wilson loop is given by \cite{LPRT}, 
\begin{equation}
\begin{split}
        \log W_{T}(r,\tau)&=g^2 C_F  \int \frac{d^{3}\vec q}{(2 \pi)^3} \frac{e^{i q_3
r}+e^{-i q_3 r}-2}{2}\Bigg(\frac{\tau}{\vec q^2+ \Pi_{L}(0,\vec q)}+\\
        &\int_{-\infty}^{\infty} \frac{dq^0}{\pi} n_{B}(q^0)[1+e^{q^0
\beta}-e^{q^0 \tau}-e^{q^0 (\beta -\tau)}]\times \\
        &\left[\left(\frac{1}{\vec q^2}-\frac{1}{(q^0)^2}\right)
\rho_{L}(q^0,\vec q) + \left(\frac{1}{q_3^2}-\frac{1}{\vec q^2}\right)
\rho_{T}(q^0,\vec q) \right]\Bigg). 
\end{split}
\label{perturb_wilson}
\end{equation}
Here $\rho_T$ and $\rho_L$ are the transverse and longitudinal components of the gluonic spectral function. This expression leads to the HTL potential in Eq.~(\ref{HTL_pot}). We observe that the real part and imaginary part originate from rather different $\tau$ dependencies of $\log W_T(r,\tau)$. The real part is determined from the linear part in $\tau$, whereas the imaginary part of the potential from the part periodic in $\tau$. 

We also observe that the potential also does not get any contribution from the  $\rho_T$ part of Eq.~(\ref{perturb_wilson}). However, the presence of this term does affect the extraction of the potential from the Euclidean lattice data.  It has been shown \cite{BR} that in leading order, Coulomb gauge fixed Wilson line correlator also reproduces the same complex potential in Eq.~(\ref{HTL_pot}), however, this correlator does not have any contribution from $\rho_T$ term. 

Motivated by this we have calculated the Coulomb gauge fixed Wilson line correlator non-perturbatively on the lattice. For the extraction of the real and imaginary part of the potential, we have then assumed the following HTL based parametrization of $W_{T}(r,\tau)$ near $\tau\sim\frac{\beta}{2}$, 

\begin{equation}
\log(W_{T}(r,\tau))=c_0 \tau +\int_{-\infty}^{\infty} d\omega \sigma(r,\omega) [\exp(\omega \tau)+\exp(\omega(\beta-\tau))].
\end{equation}
For the existence of the potential, 
\begin{equation}
	\lim_{t \rightarrow \infty} i\frac{\partial \log(W_{T}(r,t))}{\partial t}=-c_0-\lim_{t\rightarrow \infty}\int_{-\infty}^{\infty} d\omega \sigma(r,\omega) \omega [\exp(i\omega t)-\exp(\omega(\beta-i t))]=\mathrm{constant}
\end{equation}

Using the identity  $\lim_{t\rightarrow \infty}[\exp(i\omega t)-\exp(\omega(\beta-i t))]=2\pi i \omega \delta(\omega)$, we see that the above limit will exist only when  $\sigma(r,\omega)\sim \frac{1}{\omega^2}$ as $\omega \rightarrow 0$. Further using the structure of HTL perturbative results, we parametrize $\sigma(r,\omega)$ as follows,
\begin{equation}
	\sigma(r,\omega)=n_{b}(\omega)\left(\frac{c_{-1}}{\omega}+\sum_{l=0}^{\infty} c_{2l+1}\omega^{2l+1}\right)
\end{equation}

Only odd terms are present because $n_{b}(\omega)(\exp(\omega \tau)+\exp(\omega (\beta-\tau)))=-n_{b}(-\omega) (\exp(-\omega \tau)+\exp(-\omega (\beta-\tau)))$.

Using this $\sigma(\omega,r)$ we get \cite{BD},
\begin{equation}
	-\frac{\partial \log(W_{T}(r,\tau))}{\partial \tau}=-c_0+\frac{2\,\pi\,c_{-1}}{\beta} \cot\frac{\pi\tau}{\beta}+\sum_{l=0}^{\infty}c_{2l+1}G_{l}(\tau,\beta), 
\label{eff_mass}
\end{equation}
where $G(\tau,\beta)=2\frac{(2l+2)!}{\beta^{2l+3}}\left[\zeta\left(3+2l,\frac{\tau}{\beta}\right)-\zeta\left(3+2l,1-\frac{\tau}{\beta}\right)\right]$.

Using Eq.~(\ref{eff_mass}), the potential can be obtained easily, 

\begin{equation}
	V(r)=-c_0(r,T)-i\frac{2\,\pi\,c_{-1}(r,T)}{\beta}.  
\end{equation}

The real and imaginary part can then be identified as $V_{re}(r,T)=-c_0(r,T)$ and $V_{im}(r,T)=\frac{2\,\pi\,c_{-1}(r,T)}{\beta}$.
Terms inside the summation of Eq.~(\ref{eff_mass}) do not surivive in the $t\rightarrow \infty$ limit. Integrating Eq.~(\ref{eff_mass}) with the first two terms, we get the following form of the correlator around $\tau\sim\frac{\beta}{2}$,
\begin{equation}
	W_{T}(r,\tau)=A(r, T)\,\exp\left[-V_{re}(r, T)\tau-\frac{\beta V_{im}(r,T)}{\pi}\log\left(\sin\frac{\pi\tau}{\beta}\right)\right].
\label{HTL_param}
\end{equation}

\section{Lattice Spectification}
We have measured Wilson line correlators on Coulomb gauge fixed (2+1) flavor gauge field configurations generated by the HotQCD and TUMQCD collaborations with temporal extent $N_\tau=12$ and spatial extent $N_s=48$\cite{BBPVW,BPW,BBD}. The strange quark mass $m_s$ has been set to its physical value and the light quark mass to $m_l=m_s/20$. The pion mass corresponding to this parameter is 161~MeV in the continuum limit\cite{BBD}. These configurations are generated using the HISQ action for the fermionic part and Luscher-Weisz gauge action for the gluonic part. As the temporal extent is fixed, the temperature has been changed by changing the lattice spacing $a$. The lattice spacings have been fixed by the $r_1$ scale, which is defined using zero temperature static $Q\bar Q$ potential, and we use $r_1=0.3106 fm$ \cite{MILC}\cite{BPW} whenever converting to physical units.

\section{Results for the potential}
To extract the potential we fit our lattice data with the first moment $m_1(r,\tau)$ defined as follows,
\begin{equation}
	m_{1}(r,\tau)=\log\left(\frac{W_{T}(r,\tau/a)}{W_{T}(r,\tau/a+1)}\right)
\label{m1}
\end{equation}
In terms of the parametrization in Eq.~(\ref{HTL_param}) we get,
\begin{equation}
	m_{1}(r,n_\tau=\tau/a)	=V_{re}(r,T)\,a-\frac{V_{im}(r,T)a N_\tau }{\pi }\,\log\left[\frac{\sin(\pi n_\tau/N_\tau)}{\sin(\pi  (n_\tau+1)/N_\tau)}\right]
\end{equation}

\begin{figure}
\centerline{\includegraphics[width=8cm]{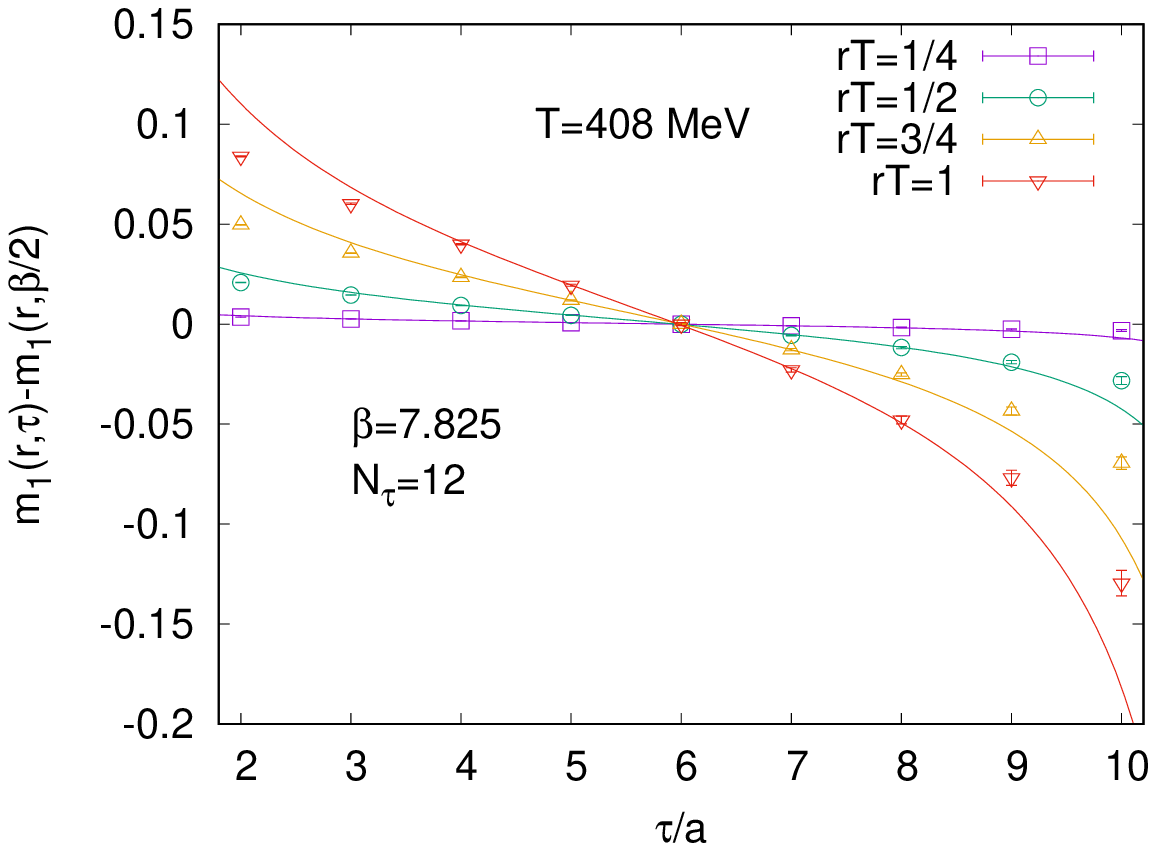}
\includegraphics[width=8cm]{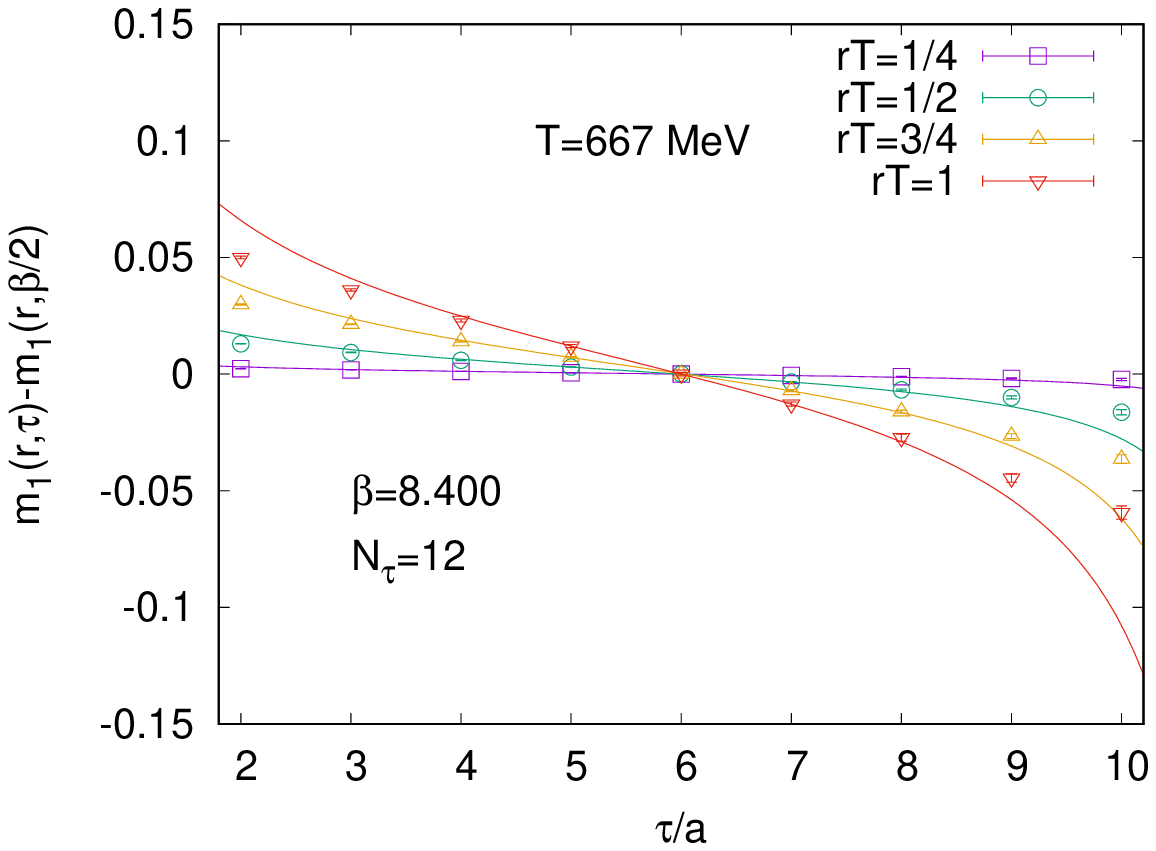}}
	\caption{Lattice data fitted with HTL motivated parametrization of Eq.~(\ref{m1}) at T=408 MeV (Left) and T=667 MeV (Right).}
\label{plateau}
\end{figure}

We have performed a two-parameter correlated fit of the equation near $\tau\sim\frac{\beta}{2}$ to extract the real and imaginary part of the potential. For lattices with $N_{\tau}=12$ we have done the fitting only with the points $\tau/a=5,6,7$, but it has been shown in the quenched approximation with large $N_{\tau}$ that reasonable data points near $\tau/a\sim N_\tau/2$ can be fitted using Eq.~(\ref{m1})\cite{BD}. The demonstration of the fit for two temperatures is shown in Fig.~\ref{plateau}. We see that the data near $\tau\sim \frac{\beta}{2}$ is well described by HTL parametrization. The smaller and higher $\tau$ range is not expected to be described by the HTL parametrization. A subtracted correlator, where the UV contaminated part is removed, has also been used to extract the real and imaginary part of the potential in \cite{BKLMPPRJ}.

\begin{figure}
\centerline{\includegraphics[width=8cm]{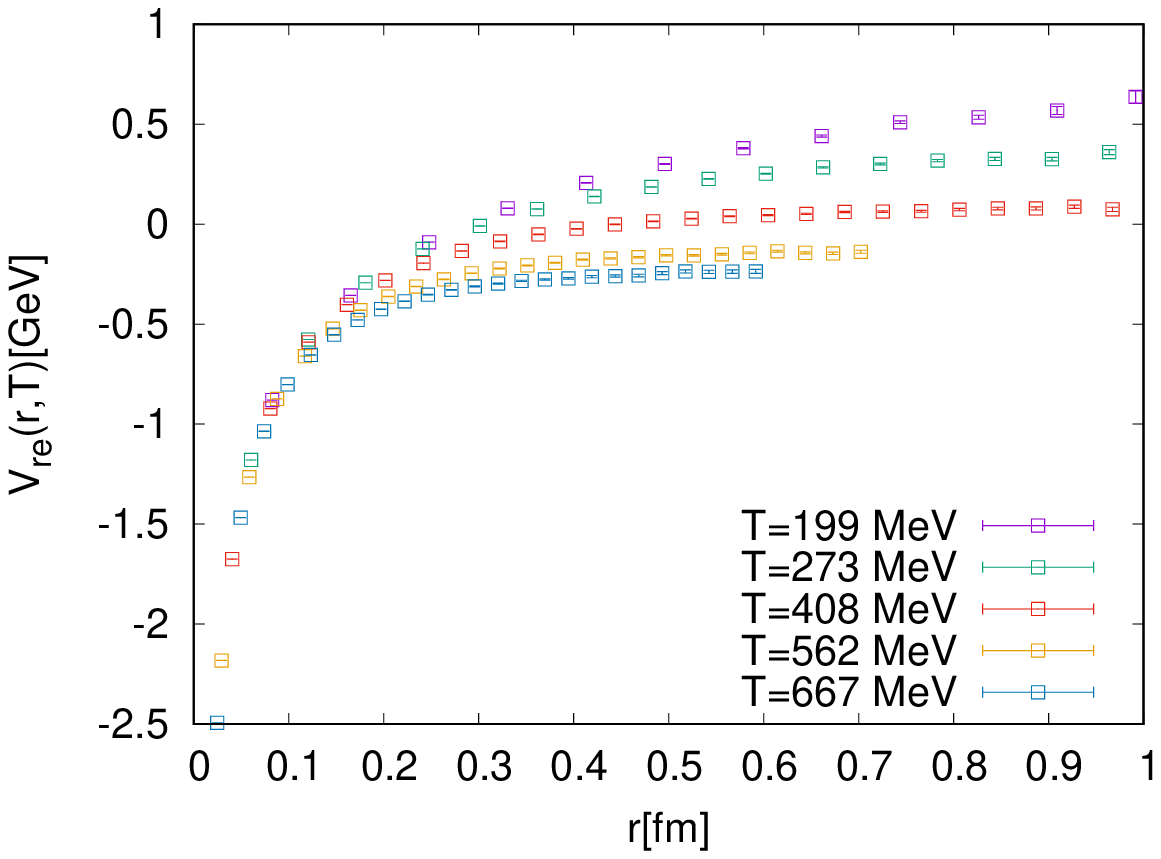}
\includegraphics[width=8cm]{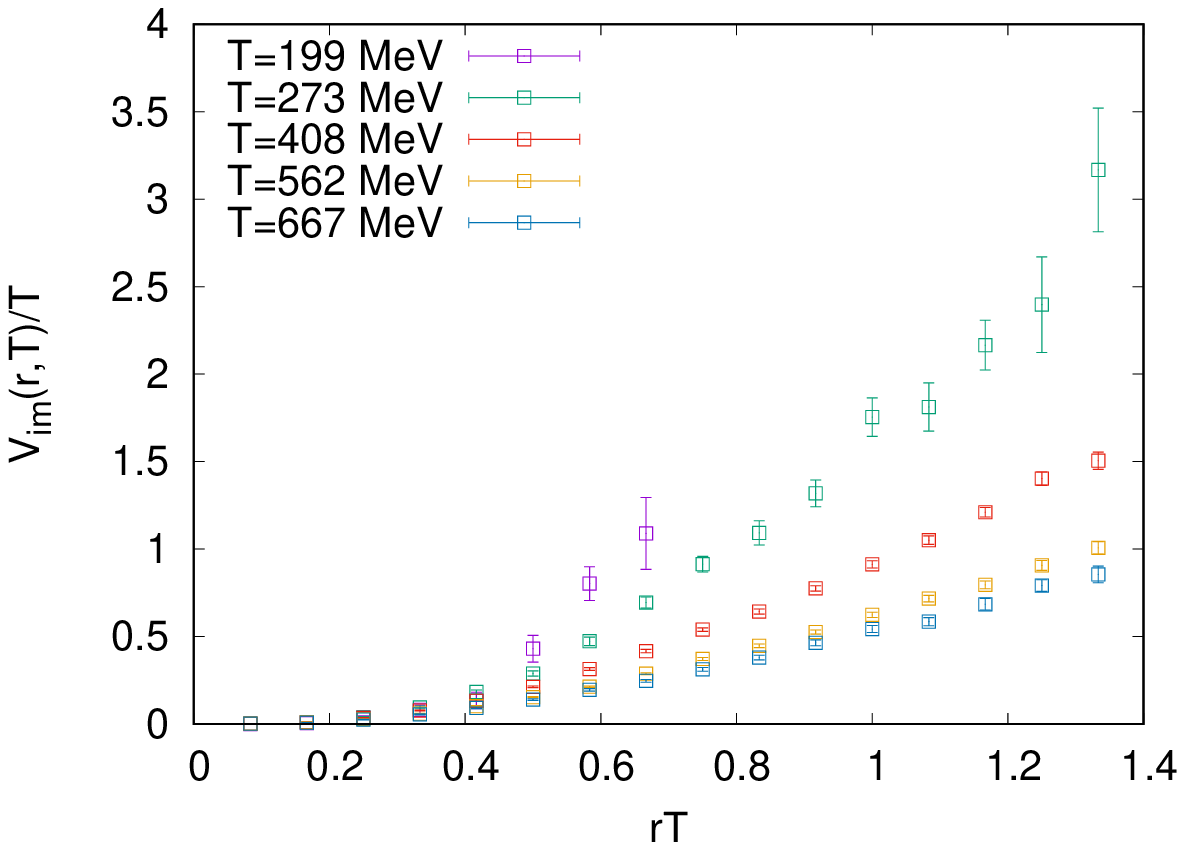}}
	\caption{The complex potential obtained from HTL parametrization at various temperature as a function of distance $r$. Real part (Left) shows medium modification above cross over temperature and Imaginary part (Right) shows non-trival temperature dependence.}
\label{potential}
\end{figure}

The real and imaginary part from this fit is shown in Fig.~\ref{potential}. We see that the real part of the potential shows medium modifications above the cross-over temperature. The large distance part of the potential shows much flatter behavior compared to the zero temperature linear behavior related to the string tension. The short-distance part of the potential shows medium modifications at a higher temperature. On the other hand, the imaginary part increases with distance at the available distances for a given temperature and approaches zero at a short distance. For a given distance the imaginary part also increases with temperature. These features of the potential are in qualitative agreement with the potential obtained in \cite{BD,BKR}. It is important to note that although the data can be fitted with the HTL parametrization, the real and imaginary parts obtained from the fits can not be quantitatively described by the HTL potential in Eq.~(\ref{HTL_pot}).

Once we have the real and imaginary part of the potential, we can compute the dominant peak of the spectral function using Eq.~(\ref{HTL_param}). The spectral function can be obtained from,
\begin{equation}
\rho(r,\omega)=\frac{1}{2\pi} \int_{-\infty}^{\infty} dt \exp(i \omega t) W_{T}(r,t)
	\label{sp1}
\end{equation}

\begin{figure}
	\centerline{\includegraphics[width=8cm]{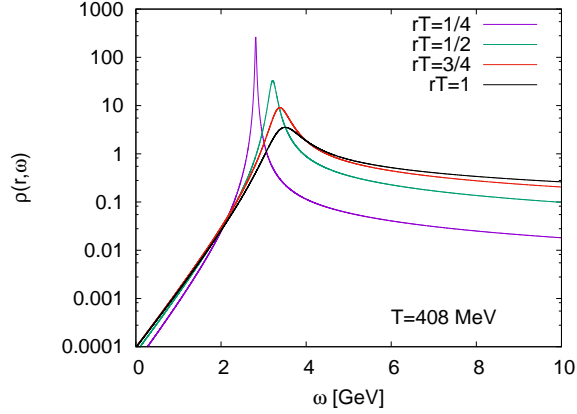}}
	\caption{Dominant peak of spectral function of Eq.~(\ref{HTL_param}).}
\label{sp_func}
\end{figure}

The spectral function is plotted in Fig.~\ref{sp_func} for a temperature of 408~MeV. We observe that the spectral function shows asymmetric structure around the peak position \cite{BR1}. The integration of  Eq.~(\ref{sp1}) can be done exactly \cite{BD1}.  
Near the peak $\rho(r,\omega)$ can be approximated by a Lorentzian,
\begin{align}
\rho(r,\omega) &\approx \sqrt{\frac{2}{\pi}} \ \frac{V_{im}(
r,T)}{(V_{re}(r,T) - \omega)^2 \; +
    \; V_{im}(r,T)^2} \qquad &{} |V_{re}(r,T) - \omega|, \;
V_{im}( r,T) \ll T \nonumber .\\
\end{align}
This is expected as we already assumed the existence of the limit in Eq.~(\ref{condition}).

Away from the peak the structure differs from a Lorentzian:
\begin{align}
\rho( r,\omega) & \sim  (\omega - V_{re}( r,T))^{- \left(1-\frac{\beta
V_{im}( r,T)}{\pi} \right)} &{} \omega -
  V_{re}( r,T) \gg T \label{asymp},\\
  & \sim  e^{- \beta (V_{re}( r,T) - \omega)} \ (V_{re}( r,T) -
  \omega)^{-\left(1-\frac{\beta V_{im}( r,T)}{\pi} \right)} &{} \omega -
V_{re}( r,T) \ll -T \nonumber.
\end{align}

The exponential suppression at the low-$\omega$ side of the peak is needed for Eq.~(\ref{cont}) to be integrable. 

\section{Comparison with singlet Free Energy}
\begin{figure}
\centerline{\includegraphics[width=8cm]{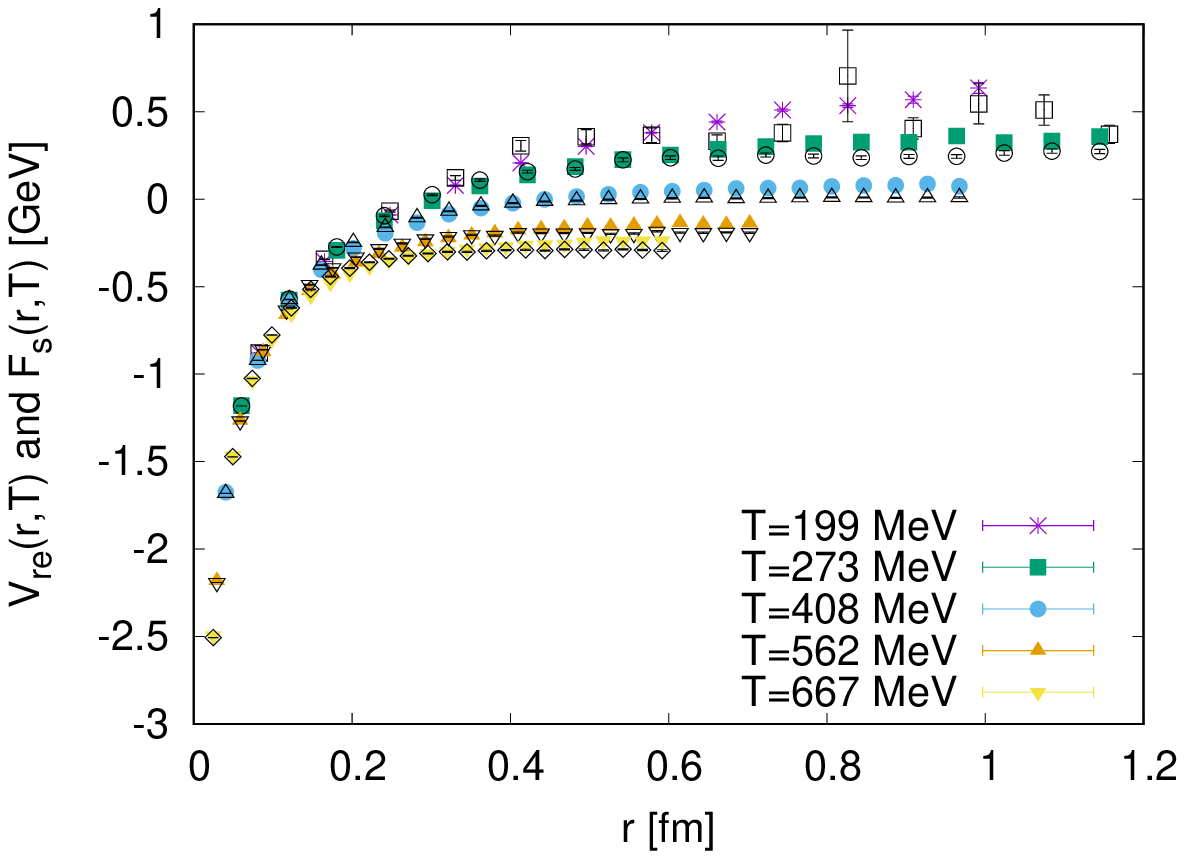}
\includegraphics[width=8cm]{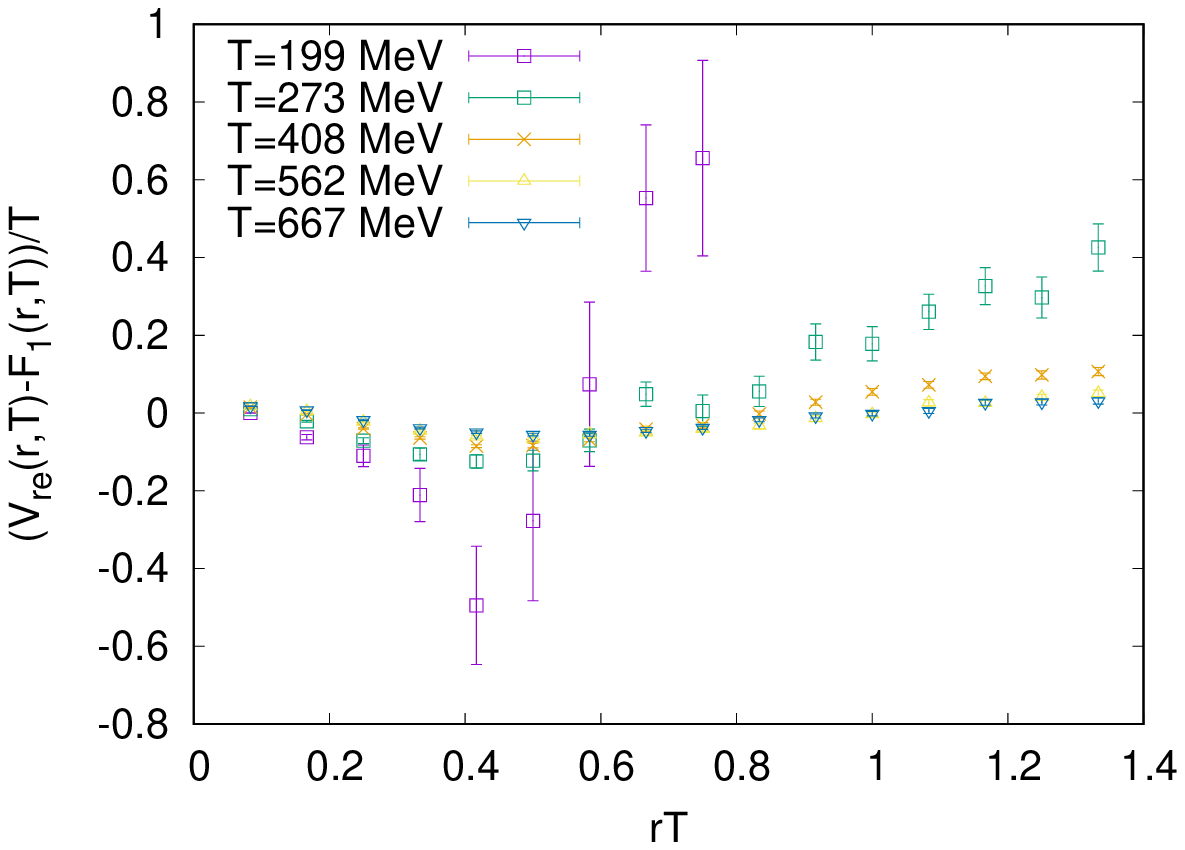}}
	\caption{(Left) The singlet free energy $F_s(r, T)$ (black points) and real part of potential $V_{re}(r, T)$ as function of $r$. (Right) $(V_{re}(r, T)-F_{s}(r, T))/T$ plotted as a function of $rT$. }
\label{free}
\end{figure}

The singlet free energy $F_s$ is another widely studied quantity \cite{KKPZ,KZ,BBPVW} at finite temperature, which is defined in Coulomb gauge and given by,
\begin{equation}
F_{s}(r,T)=-T \log(W_{T}(r,\beta)). 
\end{equation}
In leading order HTL perturbation theory, the singlet free energy is exactly equal to the real part of the potential in Eq.~(\ref{HTL_pot}).
On the left panel of Fig.~\ref{free}, the singlet free energy and the real part of the potential obtained on the lattice have been compared. 
We see that even non-perturbatively these two quantities come close to each other. However on the right panel of Fig.~\ref{free} , when we have plotted the difference $V_{re}(r,T) -F_{s}(r,T)$ we do see a finite difference between these two quantities. As we increase the temperature the difference however is becoming smaller, as at very high temperatures both of these quantities should approach to their leading order HTL value.
\section{Conclusion}
We have computed the complex static quark-antiquark potential at finite temperature using a HTL motivated approach. We have found that data near $\tau\sim\beta/2$ is consistent with the HTL parametrization. The extracted real part of the potential shows screening above the crossover temperature, whereas the imaginary part is increasing with both distance and temperature. We have also observed that the real part of the potential is very close to the singlet free energy, however, a small non-zero difference exists between them. We have also computed the structure of the dominant peak of the spectral function, which shows an asymmetric structure away from the peak and near the peak, the spectral function can be approximated by a Lorentzian.
\section{Acknowledgement}
This work is supported by the U.S. Department of Energy, Office of Science, Office
of Nuclear Physics through the (i) Contract No. DESC0012704, and (ii) Scientific Discovery through Advance Computing (SciDAC) award Computing the Properties of Matter with Leadership Computing Resources.
(iii) R.L., G.P., and A.R. acknowledge funding by the
Research Council of Norway under the FRIPRO Young
Research Talent grant 286883. (iv) J.H.W.’s research
was funded by the Deutsche Forschungsgemeinschaft
(DFG, German Research Foundation) - Projektnummer
417533893/GRK2575 “Rethinking Quantum Field Theory”. (v) D.B. and O.K. acknowledge support by the
Deutsche Forschungsgemeinschaft (DFG, German Research Foundation) through the CRC-TR 211 ’Strong interaction matter under extreme conditions’– project number 315477589 – TRR 211.

This research used awards of computer time provided by: (i) The INCITE and ALCC programs at Oak Ridge Leadership Computing Facility, a DOE Office of Science User Facility operated under Contract No. DE-AC05- 00OR22725. (ii) The National Energy Research Scientific Computing Center (NERSC), a U.S. Department of Energy Office of Science User Facility located at Lawrence Berkeley National Laboratory, operated under Contract No. DE-AC02- 05CH11231. (iii) The PRACE award on JUWELS at GCS@FZJ, Germany. (iv) The facilities of the USQCD Collaboration, which are funded by the Office of Science of the U.S. Department of Energy. (v) The UNINETT Sigma2 - the National Infrastructure for High Performance Computing and Data Storage in Norway under project NN9578K-QCDrtX "Real-time dynamics of nuclear matter under extreme conditions".

\end{document}